\documentclass[letter]{aa}
\usepackage{psfig,natbib,txfonts}

\begin{document}

\title{Properties of high-degree oscillation modes of the Sun observed with Hinode/SOT} 
\author{U.~Mitra-Kraev\inst{1} \and A.G.~Kosovichev\inst{2} \and T.~Sekii\inst{3}} 
\offprints{U.~Mitra-Kraev, \\
\email{U.MitraKraev@sheffield.ac.uk}}
\institute{University of Sheffield, Department of Applied Mathematics, Hicks
  Building, Sheffield S3 7RH, UK \and   
   W.\,W.\ Hansen Experimental Physics Laboratory, Stanford University, Stanford, CA 94305, USA \and National Astronomical Observatory of Japan, Mitaka, Tokyo 181-8588, Japan} 
\date{Received 9 November 2007 / Accepted 6 January 2008}
\titlerunning{Properties of high-degree oscillation modes of the Sun} 
\authorrunning{U.~Mitra-Kraev et al.}
\abstract{}{With the Solar Optical Telescope on Hinode, we investigate the basic properties of high-degree solar oscillations observed at two levels in the solar atmosphere, in the G-band (formed in the photosphere) and in the \ion{Ca}{ii}\,H line (chromospheric emission).}
{We analyzed the data by calculating the individual power spectra as well as the cross-spectral properties, i.e., coherence and phase shift. The observational properties are compared with a simple theoretical model, which includes the effects of correlated noise.}
{The results reveal significant frequency shifts between the \ion{Ca}{ii}\,H and G-band spectra, in particular above the acoustic cut-off frequency for pseudo-modes. The cross-spectrum phase shows peaks associated with the acoustic oscillation (p-mode) lines, and begins to increase with frequency around the acoustic cut-off. However, we find no phase shift for the (surface gravity wave) f-mode. The observed properties for the p-modes are qualitatively reproduced in a simple model with a correlated background if the correlated noise level in the \ion{Ca}{ii}\,H data is higher than in the G-band data. These results suggest that multi-wavelength observations of solar oscillations, in combination with the traditional intensity-velocity observations, may help to determine the level of the correlated background noise and to determine the type of wave excitation sources on the Sun.}{} 
\maketitle

\section{Introduction} \label{intr}
The Solar Optical Telescope \citep[SOT,][]{Tsuneta2007} on the Hinode spacecraft \citep{Kosugi2007} provides stable series of high-resolution images of the Sun in several different spectral intervals. These data offer a unique opportunity for investigating
properties of solar oscillation modes of high angular degree $\ell$ (high horizontal wave number). \citet{Sekii2007} and \citet{Nagashima2007} have used the SOT Broad-Band Filter  
Imager (BFI) data in the \ion{Ca}{ii}\,H line and the molecular G-band for high-resolution, time-distance helioseismology of supergranulation and for studying the distribution of the oscillation power in a sunspot region. \citet{Kosovichev2007}, using the \ion{Ca}{ii}\,H data, made initial observations of oscillations in the sunspot umbra, excited by a solar flare. 

A remarkable feature of the SOT Focal Plane Package \citep[FPP;][]{Tarbell2006} is its capability of providing simultaneous images in several spectral intervals with sufficiently high cadence (1 minute or shorter), thus enabling multi-wavelength, high-resolution helioseismology. Such studies are important for investigating the physical properties of solar oscillations, their excitation mechanism and interaction with the turbulent plasma with radiative losses of the photosphere and chromosphere. This knowledge is important for more accurately measuring oscillation frequencies of solar resonant modes as well as travel times for global and local helioseismology \citep{JCD2002,Thompson2004}, and also for a better understanding of how the oscillations leak into the solar atmosphere and contribute to
the chromospheric and coronal heating \citep{Erdelyi2007}.

%-----------------------------Figure Start--------------------
\begin{figure*}[t]
\centerline{
\vbox{\hbox{
\psfig{figure=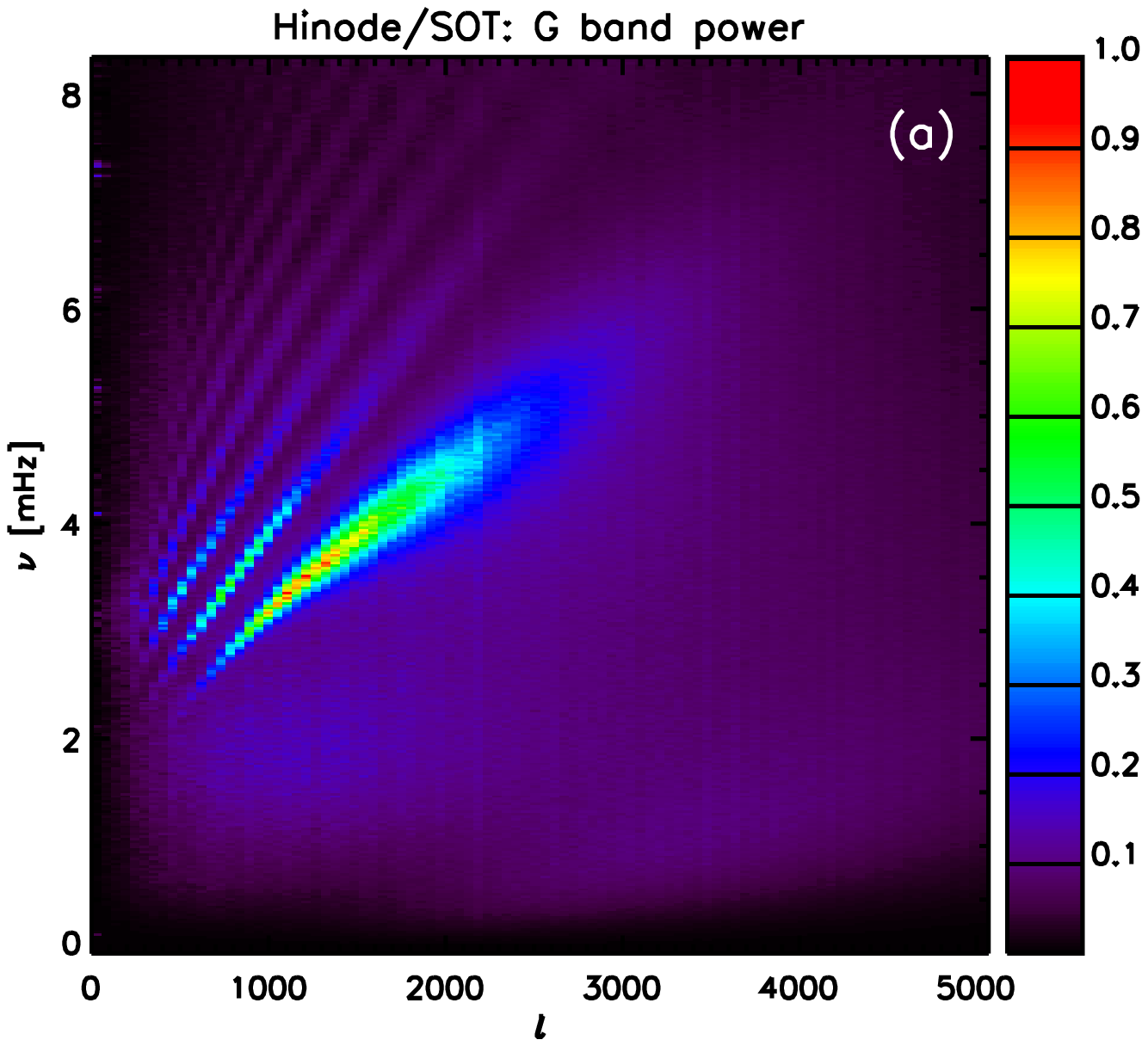,width=7.9cm}
\hspace{-2cm}
\psfig{figure=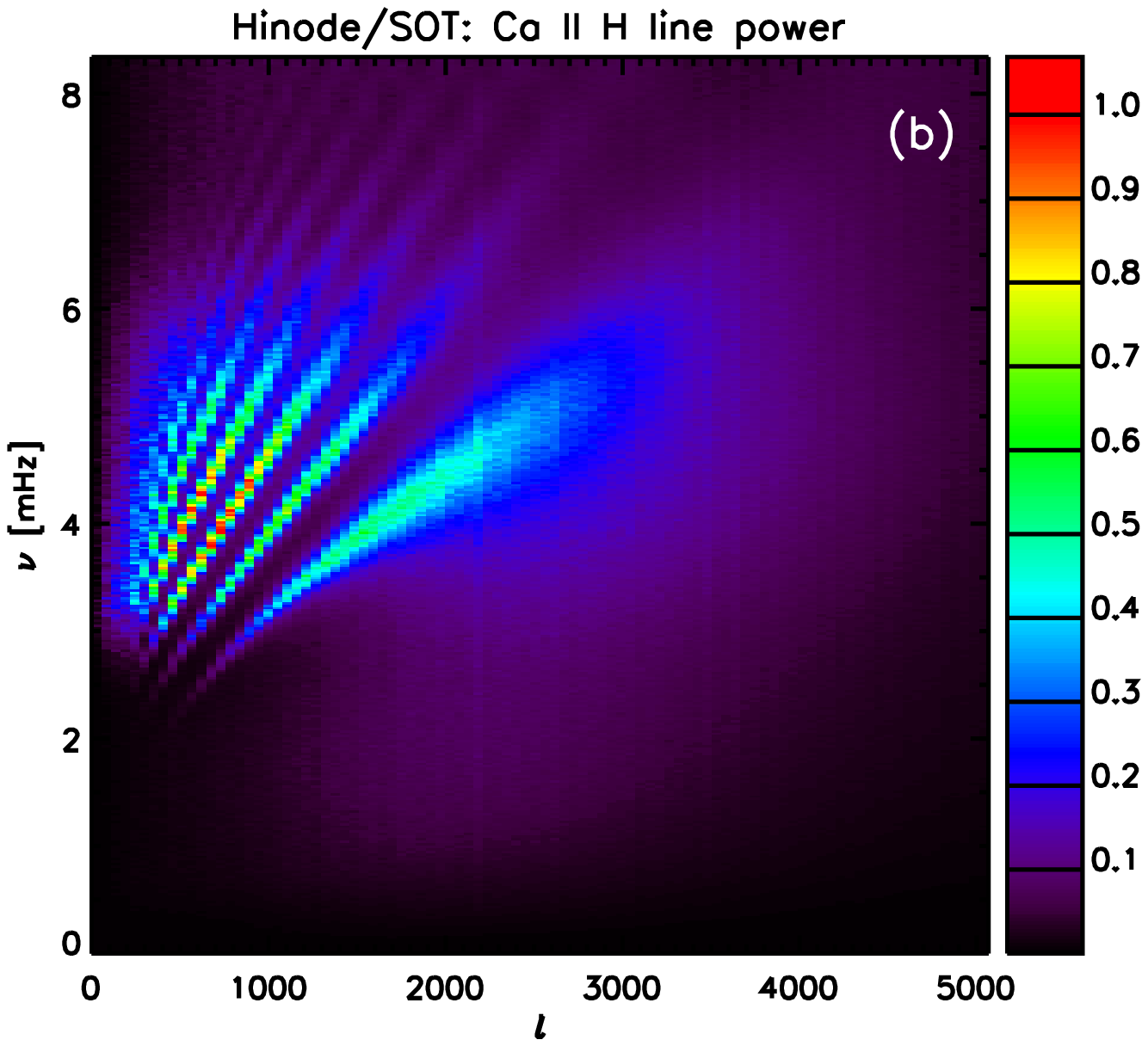,width=7.9cm}
}\vspace{.1cm}
\hbox{
\psfig{figure=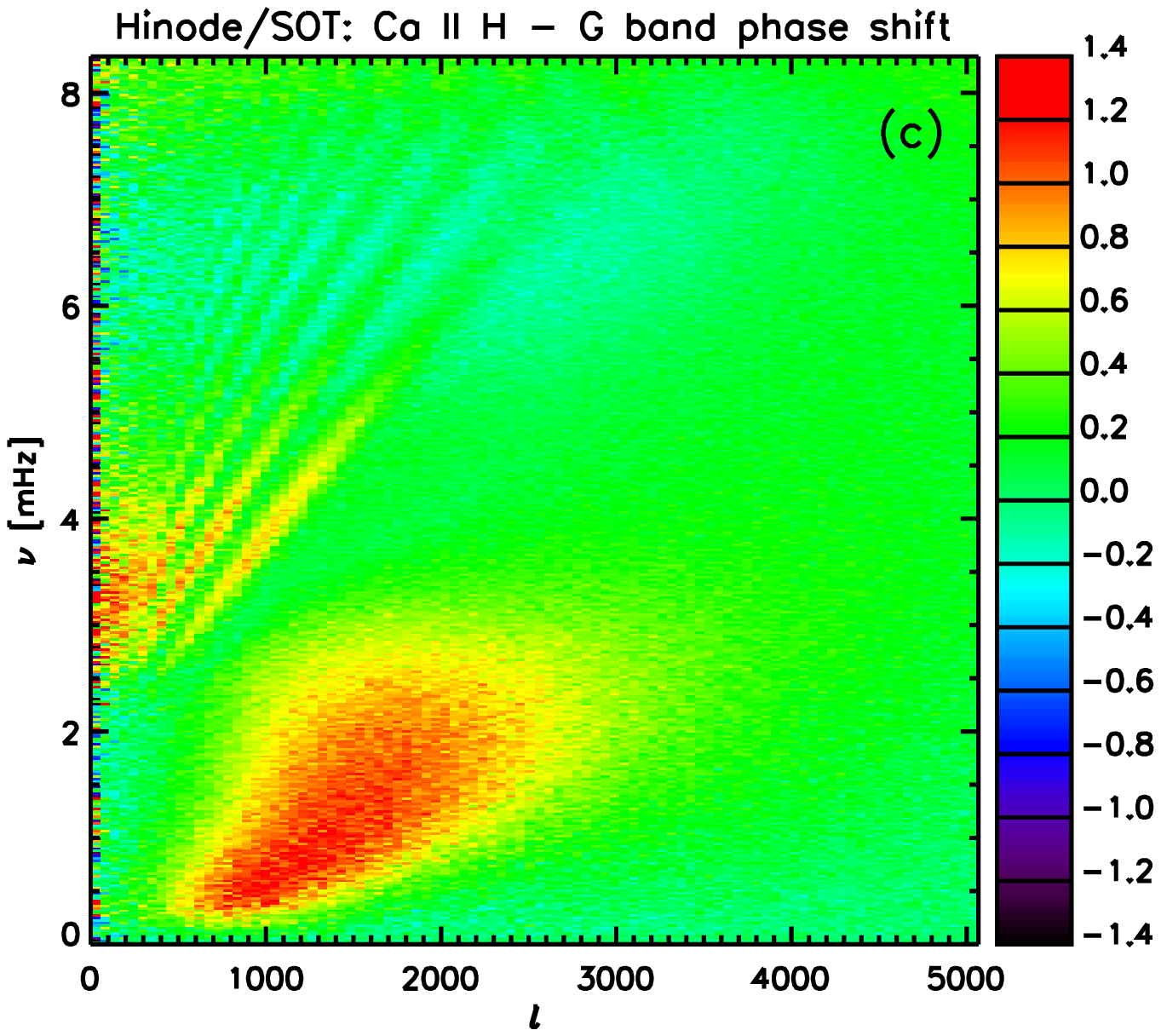,width=7.9cm}
\hspace{-2cm}
\psfig{figure=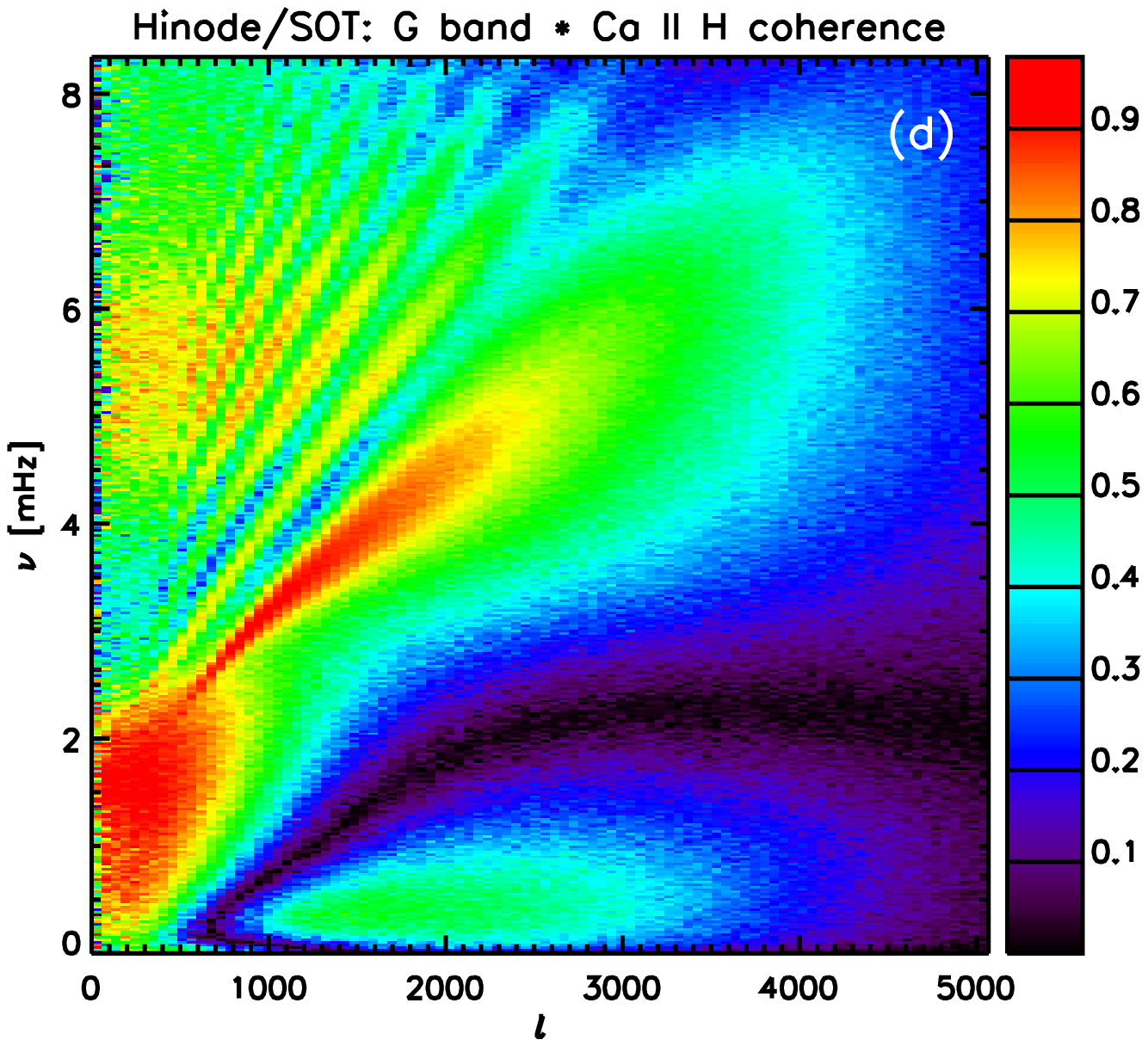,width=7.9cm}
}}}
\caption{From top left to bottom right: (a) the power spectrum from G-band observations, (b) the power spectrum from \ion{Ca}{ii}\,H line observations, (c) the phase shift between \ion{Ca}{ii}\,H and G-band (colour units are in radians), and (d) the coherence spectrum between the two lines. }
\label{fig1}
\end{figure*}
%-----------------------------Figure End---------------------------------

The goal of this paper is to investigate the spectral and cross-spectral properties of solar oscillations of high angular degree of the quiet Sun, using Hinode/SOT observations in the G-band and the \ion{Ca}{ii}\,H line, which sample the oscillation signals in the photosphere and lower chromosphere. Previously, such cross-spectral studies were carried out mostly for simultaneous intensity-velocity data at the photospheric level, using the Michelson Doppler Imager (MDI) \citep{Straus1999} onboard the Solar and Heliospheric Observatory (SOHO) and Global Oscillation Network Group (GONG) data \citep{Oliviero2000,Barban2004}. The cross-spectra reveal important information about the oscillations and the correlated and background associated with the wave excitation and propagation processes \citep[e.g.,][]{Severino2001}. 

One of the interesting properties is the asymmetry of line profiles in the oscillation power spectra \citep{Duvall1993}, which is reversed for the velocity and intensity power spectra. This is explained as an effect on the background noise correlated with the oscillation modes \citep{Nigam1998a}. The asymmetry may have a significant effect on the helioseismic diagnostics \citep[e.g.,][]{JCD1998}. The physics of the correlated background is not completely understood as the theoretical models are mostly phenomenological \citep[e.g.,][]{Nigam1998a, Rosenthal1998,Rast1999,Kumar1999,Magri2001}. The observed line asymmetry is reproduced in realistic numerical simulations \citep{Georgobiani2003}, and was linked to radiative transfer effects. Determining the level of the correlated background and relating it to the excitation mechanism (determining the depth of the excitation sources and their type) is a difficult and important problem of the Sun's oscillation physics \citep[e.g.,][]{Jefferies2003,Wachter2005}. 

In this paper, we present the cross-spectral analysis of solar oscillation observed in intensity signals at two different levels in the solar atmosphere, and show that the results are qualitatively consistent with a simple analytical model that includes the correlated background, the strength of which increases with height.

\section{Observations and results} \label{obs}
We analyze G-band 4305 and \ion{Ca}{ii}\,H line filtergram intensity data taken by Hinode/SOT. The observation started at 2007-01-01T16:06:29 and lasted for 11.87 hours. The cadence is 60$\pm$1\,s, with the \ion{Ca}{ii}\,H line data trailing the G-band data by 10\,s. The observation was taken on the quiet Sun. Figure~1 of \citet{Sekii2007} shows a snapshot of a part of the observed field in the photosphere (G-band, top panel) and chromosphere 
(\ion{Ca}{ii}\,H, bottom panel). We used the entire observed field-of-view of the camera, 160\,Mm$\times$80\,Mm, or 223.15$\times$111.575\,arcsec.

For each waveband, we calculated the Fourier and power spectrum from the time-difference images, having first binned the images by a factor of four. We also obtain the phase shift between \ion{Ca}{ii}\,H and G-band as well as their coherence. Given the 3D Fourier spectra $\hat{g}(k_x,k_y,\nu)\equiv r_g {\rm e}^{i\phi_g}$ and $\hat{h}(k_x,k_y,\nu)\equiv r_h {\rm e}^{i\phi_h}$ of two data cubes $g(x,y,t)$ (e.g., G-band) and $h(x,y,t)$ (e.g., \ion{Ca}{ii}\,H), we define the inner product:
\[
< \hat{g},\hat{h} >_{\ell\nu} = 
	\sum_{(\ell-\Delta\ell)^2\le (k_x^2+k_y^2)R_\odot^2 < (\ell+\Delta\ell)^2}
		\hat{g}^*(k_x,k_y,\nu)\ \hat{h}(k_x,k_y,\nu).
\] 
The 2D power spectra are then defined as $P_g={<\hat{g},\hat{g}>}$ and $P_h={<\hat{h},\hat{h}>}$, respectively, while the coherence spectrum COH and the phase shift $\Delta\Phi$ are given by 
\begin{eqnarray*} 
{\rm COH} &=& \frac{\left| <\hat{g},\hat{h}> \right|}{\sqrt{P_gP_h}}\\ 
	\Delta\Phi(\ell,\nu) &=& \frac{1}{N} 
		\sum_{(\ell-\Delta\ell)^2\le (k_x^2+k_y^2)R_\odot^2 < (\ell+\Delta\ell)^2}
			\phi_h(k_x,k_y,\nu)-\phi_g(k_x,k_y,\nu),  
\end{eqnarray*}  
where $N$ is the number of elements over which the sum is taken. 

Being the normalized vector product, the coherence measures the angle between the two ``vectors'' $\hat{g}$ and $\hat{h}$ -- the value is 1 if the vectors are parallel, and 0 if they are perpendicular -- and is a quantitative measure of the similarity of the two signals. It can be shown that the horizontal phase shift between two waves at different heights (photosphere and chromosphere here) is proportional to ${\rm Re}(k_z)\cdot \Delta z$, where $\Delta z$ is the height difference \citep{DeubnerFleck1989}, if the wave is propagating in the $z$ direction.

%-----------------------------Figure Start--------------------
\begin{figure*}[t]
\centerline{
\psfig{figure=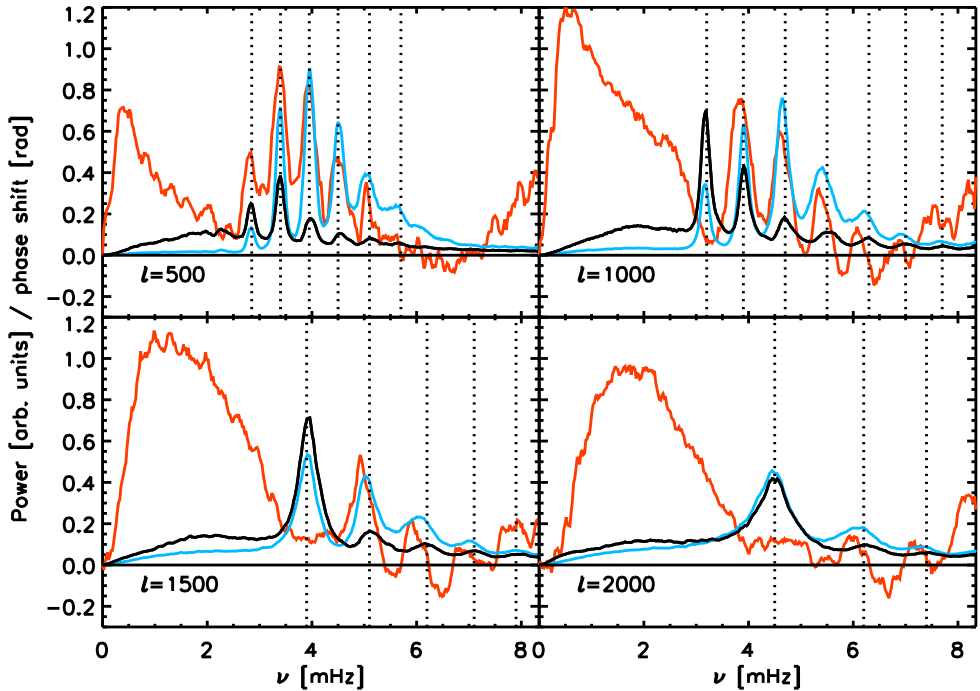,width=13.5cm}}
\caption{Cuts through the G-band (black line) and \ion{Ca}{ii}\,H (blue line) power spectra and their phase shift (red line) at different $l$ values. The vertical dotted lines are plotted at f- and p-mode peak values of the G-band power spectrum.} 
\label{fig2}
\end{figure*}
%-----------------------------Figure End---------------------------------
Figure~\ref{fig1} shows the two power spectra, the phase shift and coherence spectrum. Figure~\ref{fig2} shows four cuts through the two power spectra (black line: G-band, blue line: \ion{Ca}{ii}\,H) and the phase shift (red line) taken at spherical harmonic degrees $l=500,1000,1500$, and $2000$. The power spectra reveal that the solar oscillation spectrum is extended to the angular degree as high as 4000, and also above 8.3\,mHz, the Nyquist frequency  of the observations. The f-mode ridge becomes very broad at high-$\ell$,  \citet{Duvall1998} related to scattering on turbulence. 

Below the acoustic cut-off frequency ($\sim5.2$\,mHz), the p-mode spectral lines are  
asymmetric with the typical asymmetry of intensity oscillations \citep{Duvall1993,Nigam1998a}. Above the acoustic cut-off, the ridges in the power spectra correspond to pseudo-modes arising from the source resonance \citep{Kumar1991}, and their position depends on the source location below the solar surface and the correlated background level \citep{Nigam1998a}. The power spectra profiles in Fig.~\ref{fig2} show that the peaks 
in the \ion{Ca}{ii}\,H spectra are displaced to lower frequencies relative to the G-band peaks. This indicates that the correlated background levels are different in photospheric and chromospheric intensity oscillations. In the next section, we show that this can be modeled by assuming a higher level in the \ion{Ca}{ii}\,H (chromospheric) oscillations.

Perhaps the most unexpected result of this analysis is that the phase shift in the cross-spectrum has strong peaks along the p-mode ridges below the acoustic cut-off frequency, but vanishes in the f-mode ridge (except for  $\ell=500$, which shows up in Fig.~\ref{fig2}). Assuming that the oscillation modes represent standing waves below the cut-off frequency one would expect that the phase shift is also zero for the p-modes. Of course, the correlated background changes this \citep{Nigam1999}, and our model confirms this fact, but why the p- and f-modes  are affected differently is a puzzle. Another explanation could be the possible upward propagation of the p modes. The surface-gravity (f) mode on the other hand does not show any phase shift because this mode only propagates horizontally and is incompressible. 
There seems to be a small frequency shift with the opposite sign between the p-mode ridges above the acoustic cut-off frequency. All in all, more work needs to be done to correctly account for the correlated noise, non-LTE effects, radiative transfer, and other physics.

In the G-band and the coherence spectra, we notice a broad inclined structure below the f-mode ridge and $\ell$ below 2000. It remains to be seen whether this may correspond to atmospheric g-modes or represent some properties of solar granulations. In the \ion{Ca}{ii}\,H spectrum, there is a depression in this region.

\section{Theoretical Interpretation}\label{sect:model}
To check whether the observational results are consistent with the current theoretical ideas of the role of the correlated background, we used a simple analytical model in which instead of the complicated acoustic potential \citep{Deubner1984} a one-dimensional square potential well was used \citep{Abrams1996}. This model had been used by \citet{Kumar1999} for explaining the properties of asymmetry reversal, and by \citet{Nigam1998b} for deriving  a fitting formula for asymmetrical line profiles in the oscillation power spectra.

We used a solution for a simple $\delta$-function source, given by Eq.~(3) in \citet{Kumar1999}, and added a frequency-independent correlated signal to this solution. We adopted the following parameters: the acoustic cut-off frequency $\alpha/2\pi=5.2$~mHz; the damping parameter $\gamma=5\times 10^{-4}$~s; the sound travel times from the lower boundary $a=600$~s; and the sound travel time from the lower boundary to the source location $r_s=540$~s. The travel times to the observing levels are: 630~s for the photospheric (G-band) signal, and 660~s for the chromospheric ({\ion{Ca}{ii}\,H). The correlated background level was adjusted to get the power spectral lines and the cross-spectrum phase qualitatively similar to the observations. These values in the non-dimensional units of Eq.~(3) of \citet{Kumar1999} are: 20 for the G-band oscillation model and 25 for the \ion{Ca}{ii}\,H model. The higher level of the correlated background is necessary to explain the shift in the locations of the peak maxima (corresponding to the pseudo-modes) above the acoustic cut-off frequency. This may correspond to the higher noise level in the \ion{Ca}{ii}\,H power spectrum (Fig.~\ref{fig2}), however, we made no attempt to separate the correlated and uncorrelated parts in the data. This requires quantitative modeling of the oscillation signal. 

%-----------------------------Figure Start--------------------
\begin{figure}[ht]
\centerline{
\psfig{figure=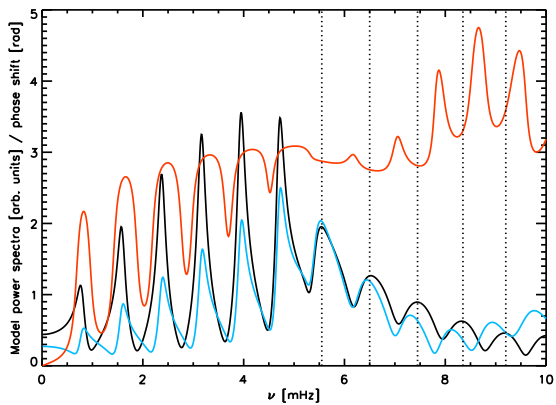,width=8cm}}
\caption{Power spectra and the phase cross-spectrum for a simple oscillation model (Sec.~\ref{sect:model}), which qualitatively represents the oscillation signals at two heights with different correlation background levels: in the photosphere (black curve) and in the chromosphere (blue curve). The red curve shows the phase shifted by $\pi$. This cut is at no specific wave number, but reflects the general behavior of the p-mode peaks above and below the acoustic cut-off frequency.}
\label{fig3}
\end{figure}
%-----------------------------Figure End---------------------------------
The model results in Fig.~\ref{fig3} show surprisingly good qualitative agreement with the observations, confirming that the correlated background is an essential part of the observed intensity signals. However, despite its similar shape, the cross-spectrum phase difference has a constant offset of about $-\pi$ relative to the observations. We do not yet have an explanation for this. 

Of course, this simple model just gives a reasonable interpretation of the power spectra of the intensity data, obtained from Hinode. More precise modeling, such as developed by \citet{Wachter2005}, is required to explain the physics of the observed oscillations in
detail. However, it is reasonable to expect that the cross-spectra of solar oscillations, measured at different heights of the solar atmosphere in combination with the intensity-velocity spectra, will help to disentangle the contributions of the correlated and uncorrelated parts, and get better estimates for the acoustic source depth and type.

\section{Discussion and conclusions} \label{disc}

The initial high-resolution observations of solar oscillations with Hinode, simultaneously in two spectral intervals, reveal new interesting properties. The mode ridges are extended in both the high angular degree (up to $\ell=4000$) and high-frequency regions (above the Nyquist frequency of 8.33~mHz). For $\ell > 2500$, the frequencies of all modes become higher than the acoustic cut-off frequency, but the oscillation power is still concentrated along the mode ridges. There is an indication that the p-mode power in \ion{Ca}{ii}\,H is higher at high frequencies than in the G-band spectrum. This is consistent with traveling wave properties as expected from solar oscillation theory. 

The cross-spectrum phase difference between the G-band and \ion{Ca}{ii}\,H oscillations shows rather unexpected properties. It is striking that the phase difference is quite large along the p-mode ridges below the acoustic cut-off frequency, but is completely absent for the f-mode. The naive standing wave picture for the resonant modes trapped below the acoustic cut-off frequency suggests that the phase difference should also be close to zero along the p-mode ridges. We do not have any explanation for the different phase behaviors of f- and p-modes. Above the acoustic cut-off frequency, the phase difference for the p-modes increases with frequency, which is also consistent with the picture of traveling waves.

Some basic properties of these observations can be qualitatively explained in a simple analytical model of solar oscillations, excited by a localized source \citep{Abrams1996}, if an extra term corresponding to the so-called correlated background is added to the wave signal \citep{Nigam1998a}. The correlated background affects the asymmetry of the modal lines below the acoustic cut-off frequency, and shifts the frequencies of the pseudo-modes above the cut-off. To explain this frequency shift, the relative level of the correlated background
in the \ion{Ca}{ii}\,H data should be higher than in the G-band signal. The physics of the correlated background is not yet fully understood, but it plays an important role in defining the line profile of the resonant modes, and the positions of the pseudo-modes. There have been many attempts to measure the level of the correlated background (and disentangle
it from the uncorrelated noise) by fitting the intensity-velocity cross-spectra obtained from the same spectral line. Hinode/SOT multi-wavelength observations open a new opportunity for studying this important aspect of the physics of solar oscillations.

\begin{acknowledgements}
Hinode is a Japanese mission developed and launched by ISAS/JAXA, with NAOJ as domestic partner and NASA and STFC (UK) as international partners. It is operated by these agencies in co-operation with ESA and NSC (Norway). This work was partly carried out at the NAOJ Hinode Science Center, which is supported by the Grant-in-Aid for Creative Scientific Research ``The
Basic Study of Space Weather Prediction'' from MEXT, Japan (Head Investigator: K. Shibata), generous donations from Sun Microsystems, and NAOJ internal funding.
\end{acknowledgements}

%\appendix
%\input{appendix}

\bibliographystyle{aa}
\bibliography{ref}

\end{document}